\documentclass[aps,twocolumn,showpacs]{revtex4}
\usepackage{amsmath,amssymb}
\usepackage{graphicx}
\newcommand{\calH}{\mathcal{H}}
\newcommand{\tilH}{\tilde H}
\newcommand{\tilE}{\tilde E}
\newcommand{\vv}[1]{\boldsymbol #1}
\newcommand{\bra}[1]{\langle\hspace{0.45pt}{#1}\vert}
\newcommand{\ket}[1]{\vert{#1}\hspace{0.45pt}\rangle}

\begin{document}
\title{Exact spin-cluster ground states in a mixed diamond chain}
\author{Ken'ichi Takano$^{1}$, Hidenori Suzuki$^{1}$, and Kazuo Hida$^{2}$} 
\affiliation{$^{1}$Toyota Technological Institute, Tenpaku-ku, Nagoya 468-8511, Japan\\
$^{2}$Division of Material Science, 
Graduate School of Science and Engineering, \\ 
Saitama University, 
Saitama, Saitama 338-8570, Japan}
\date{2009.3.31}
\begin{abstract}
The mixed diamond chain is a frustrated Heisenberg chain composed of successive diamond-shaped units with two kinds of spins of magnitudes $S$ and $S/2$ ($S$: integer).  
Ratio $\lambda$ of two exchange parameters controls the strength of frustration.  
With varying $\lambda$, the Haldane state and several spin cluster states appear as the ground state.  
A spin cluster state is a tensor product of exact local eigenstates of cluster spins.  
We prove that a spin cluster state is the ground state in a finite interval of $\lambda$.  
For $S = 1$, we numerically determine the total phase diagram consisting of five phases. 
\end{abstract}
\maketitle
%--------------------------------------------------------------------

Effects of frustration under strong quantum fluctuation 
in low-dimensional quantum spin systems have been 
of great  interest in condensed matter physics. 
Among them, the diamond chain consisting of successive 
diamond-shaped units has been attracting remarked attention. 

It has been rigorously shown 
that a  ground state of the uniform spin diamond chain (UDC) 
is a singlet spin cluster state with spontaneous breakdown 
of translational symmetry \cite{TKS}. 
The spin cluster state is an essentially quantum state 
described as a tensor product of local singlet states 
each consisting of several spins. 
For a usual gapped spin liquid, such a state is an approximation 
or is exact only at a single parameter value. 
However, in the UDC, the spins in each cluster forms an exact local singlet, and the exact solution is realized 
over a finite parameter range. 

The distorted version of this model has been also investigated 
theoretically \cite{OTTK,OTK,sano,honecker}. 
Since its discovery in a natural mineral azurite \cite{kiku,ohta}, 
this system is attracting a renewed interest. 
Also materials in a unfrustrated parameter region are 
found \cite{izuoka,uematsu}.

%-->-->-->-->-->-->-->-->-->-->-->-->-->-->-->-->
%Fig. 1
\begin{figure}[b] %[!btp] [H]
\begin{center}\leavevmode
\includegraphics[width=0.75\linewidth]{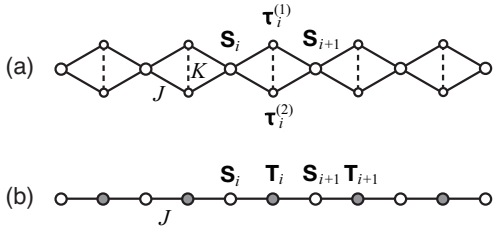}
\end{center}
\vskip -0.5cm
\caption{
(a) The MDC, where spin $\mathbf{S}_{i}$ has a magnitude 
$S$ (a positive integer) and spin $\vv{\tau}_{i}^{(\alpha)}$ 
$(\alpha = 1, 2)$ has $\tau = \frac12 S$. 
The solid (dashed) line represents exchange parameter $J$ ($K$). 
(b) The equivalent lattice to the lattice in (a) 
for each set of $\{ T_i \}$, where 
$T_i$ is the magnitude of ${\bf T}_{i} =$ 
$\vv{\tau}_{i}^{(1)}$ + $\vv{\tau}_{i}^{(2)}$. 
} 
\label{lattice}
\end{figure}
%--<--<--<--<--<--<--<--<--<--<--<--<--<--<--<--<

In this paper, we report rigorous ground states 
including spin clusters which are not singlet. 
We found the cluster states for the mixed diamond chain (MDC) 
which has the same lattice structure as the UDC as 
shown in Fig.~\ref{lattice}(a). 
Unlike the UDC, the MDC consists of 
two kinds of spins with magnitudes $S$ and $\frac{1}{2} S$ 
($S$: a positive integer). 
The ratio of two exchange constants in this model is 
a parameter representing the strength of the frustration. 
For a weak frustration case, 
the ground state of the MDC is the Haldane state~\cite{haldane}, 
while that of the UDC is ferrimagnetic. 
An interest is how the ground state changes from 
the Haldane state when frustration increases. 
We show that the MDC undergoes a series of characteristic 
quantum phase transitions with increasing frustration. 
A typical ground state is of a tensor product form of 
finite-length triplet clusters and dimers in an alternating array; 
thus the translational symmetry is spontaneously broken.
Rigorous results on the phase diagram 
are presented for general values of $S$. 
In particular, a tensor-product state is proven to be 
the ground state for a finite parameter range. 
For $S=1$, the whole phase diagram is numerically determined. 

The cluster ground states of the MDC have a macroscopic 
degeneracy, because each non-singlet cluster degenerates 
with respect to the spin direction. 
Such macroscopic degeneracy can give birth to 
a rich variety of quantum phenomena 
in the presence of perturbation. 
Even within our preliminary studies,  quantum ferrimagnetism, 
unusual Haldane phases with spontaneous breakdown of 
translational symmetry, and infinite series of quantum phase 
transitions are found in the presence of lattice distortion.

The MDC consisting of $N$ diamond units 
is described by the Hamiltonian: 
%------------------------------------------------------------
\begin{align}
\label{hamiltonian} %(1)
H_N &=\sum_{i=1}^{N} \calH_{i} , \\
\label{sub_ham} %(2)
\calH_{i} &= J({\bf S}_{i}+{\bf S}_{i+1})\cdot 
(\vv{\tau}_{i}^{(1)}+\vv{\tau}_{i}^{(2)})
+K\vv{\tau}_{i}^{(1)}\cdot \vv{\tau}_{i}^{(2)} . 
\end{align}
%------------------------------------------------------------
In the $i$th unit cell, 
${\bf S}_{i}$ is a spin whose magnitude $S$ is an integer, 
while $\vv{\tau}_{i}^{(\alpha)}$ $(\alpha = 1, 2)$ is a spin 
of magnitude $\tau = \frac{1}{2} S$. 
Below we sometimes use the notations, 
${\tilde {\bf S}_{i}}$ $\equiv$ ${\bf S}_{i}$ + ${\bf S}_{i+1}$ 
and ${\bf T}_{i}$ $\equiv$ $\vv{\tau}_{i}^{(1)}$ + 
$\vv{\tau}_{i}^{(2)}$. 
We assume that $J > 0$. 
Then the ratio $\lambda \equiv K/J$ represents the strength 
of frustration if $K > 0$. 

We first examine the classical limit of $S\rightarrow\infty$ 
with finite $JS^{2}$. 
(i) For $\lambda \le 2$, we write the classical version of 
Eq. (\ref{sub_ham}) as 
$\calH_{i}^{\mathrm{cl}} = \frac{1}{4} J 
[ (2{\bf T}_{i}+{\bf{\tilde S}}_{i})^{2} 
- 2(2 -\lambda) {\bf T}_{i}^{2} 
- {\bf{\tilde S}}_{i}^{2} - \lambda S^2 ]$. 
This is minimized  
if $|{\bf T}_{i}|=2\tau =S$, $|{\bf \tilde S}_{i}|=2S$ 
and $|{\bf T}_{i}+\frac{1}{2}{\bf\tilde S}_{i}|=0$. 
Then all the ${\bf S}_{i}$'s ($\vv{\tau}_{i}^{(\alpha)}$'s) 
in the chain are aligned parallel (antiparallel) to a fixed axis, 
and the ground state is antiferromagnetic. 
This ground state is elastic, since any local modification of 
the spin configuration increases the energy. 
(ii) For $\lambda > 2$, we use the expression 
$\calH_{i}^{\mathrm{cl}} = (J/4\lambda) 
[2(\lambda{\bf T}_{i}+{\bf\tilde S}_{i})^{2} 
- 2{\bf\tilde S}_{i}^{2} - \lambda^2 S^2]$. 
This is minimized if $|{\bf \tilde S}_{i}|=2S$ 
and $|{\bf T}_{i}+{\bf\tilde S}_{i}/\lambda |=0$. 
Hence ${\bf S}_{i}$ and ${\bf S}_{i+1}$ are parallel, and 
$\vv{\tau}_{i}^{(1)}$ and $\vv{\tau}_{i}^{(2)}$ 
form a triangle with ${\bf \tilde S}_{i}/\lambda$. 
$\vv{\tau}_{i}^{(1)}$ and $\vv{\tau}_{i}^{(2)}$ 
may be rotated about 
the axis of ${\bf S}_{i}$ and ${\bf S}_{i+1}$ 
without raising the energy. 
Then all the ${\bf S}_{i}$'s in the chain are aligned 
parallel to a fixed axis, and the arbitrariness of the local rotation 
of $\vv{\tau}_{i}^{(1)}$ and $\vv{\tau}_{i}^{(2)}$ is not obstructed. 
Thus the ground state is ferrimagnetic 
with magnetization $(1 - 2/\lambda)SN$. 

Returning to the quantum case with general $S$, the MDC 
has a series of conservation laws  
%------------------------------------------------------------
\begin{align}
[{\bf T}_{i}^{2}, H_N]=0 \quad (i = 1, 2, \cdots , N);  %(3)
\end{align}
%------------------------------------------------------------
i.~e. $T_{i}$ ($= 0, 1,\cdots, S$) of 
${\bf T}_{i}^{2}=T_{i}(T_{i}+1)$ for each $i$ 
is a good quantum number. 
With a fixed set $\{ T_{i}\}$,
the original problem of 3$N$ spins reduces to a problem of a linear chain with 2$N$ spins, where the ($2i-1$)th site is occupied by the spin ${\bf S}_{i}$ and the 2$i$th by ${\bf T}_{i}$. 
The energy of $K$-bonds is determined solely by $\{ T_{i}\}$. 
In particular, if $T_{i}=0$, then there is no interaction between 
the left and right sides of ${\bf T}_{i}$ 
and the whole lattice is decoupled. 
We denote the lowest energy of $H_N$ for a fixed set 
$\{ T_i \}$ as $E_N (\{ T_i \}, \lambda)$, and sometimes 
drop $\{ T_i \}$ and/or $\lambda$ in it. 
Hereafter we use an energy unit of $J=1$. 

There is an equivalent lattice to the MDC 
for each fixed set of $\{ T_{i}\}$, because 
${\bf T}_{i}^{2}$s are conserved. 
The equivalent lattice for set $\{ T_{i}\}$ is 
a nearest-neighbor antiferromagnetic linear spin chain 
on $2N+1$ sites. 
The spin magnitudes are $S$ on all odd sites and 
$T_{i}$ on $2i$th site for all $i$, 
as seen in Fig. \ref{lattice}(b). 
The Hamiltonian is 
%------------------------------------------------------------
\begin{align}
\label{equiv_Ham} %(4)
\tilH_N &= \sum_{i=1}^{N} 
({\bf S}_{i}+{\bf S}_{i+1})\cdot {\bf T}_{i} . 
\end{align}
%------------------------------------------------------------
$\tilH_N$ is related to $H_N$ as $H_N =$
$\tilH_N + \frac{1}{2} \lambda \sum_{i=1}^N [ T_i (T_i + 1) 
- \frac{1}{2} S(S+2)]$. 
The ground state energy of $\tilH_N$ is denoted by 
$\tilE_{N}(\{ T_i \})$, or simply $\tilE_{N}$. 

We begin with analyzing eigenstates of 
an isolated diamond unit described by $\calH_{i}$. 
This is realized in the total system, 
if $T_{i-1}=T_{i+1}=0$. 
Using the expression, 
%------------------------------------------------------------
\begin{align}
\label{sub_ham_1} %(5)
\calH_{i} = \frac{1}{2} 
 \{ ({\bf T}_{i}+{\bf{\tilde S}}_{i})^{2} 
+ (\lambda - 1) {\bf T}_{i}^{2} 
- {\bf{\tilde S}}_{i}^{2} \} 
 - \frac{\lambda}{4} S (S + 2) , 
\end{align}
%------------------------------------------------------------
the lowest energy of $\calH_{i}$ for a given $T_{i}$ is 
%------------------------------------------------------------
\begin{align}
\label{energy1} %(6)
E_{1}(T_{i}) = \frac12 
T_{i}(T_{i}+1) [\lambda - \Lambda(T_{i})] 
- \frac14 \lambda S(S + 2), 
\end{align}
%------------------------------------------------------------
where $\Lambda(L) \equiv 2(2S + 1)/(L + 1)$ 
for nonnegative integer $L$. 
Then the total spin of the diamond unit is $2S-T_{i}$. 
For the whole spin chain, we have the following lemma: 

%=============================
{\bf Lemma 1.} 
If $L \ge 1$ and $\lambda > \Lambda(L)$, 
then $T_{i}\ne L$ for any $i$
in the ground state of the MDC. 
%=============================

{\bf Proof.} 
The total Hamiltonian is divided as $H = \calH_{m} + H'$, 
where $N$ of $H_N$ is omitted and 
$H'$ is the sum of $\calH_{i}$'s with $i\ne m$. 
We take a state 
$|\Psi_{0} \rangle$ = $|0 _m \rangle \otimes |\Psi' \rangle$ 
where $|0 _m \rangle$ is the singlet wave function 
of ${\bf T}_{m}$ and $|\Psi' \rangle$ 
is the lowest-energy state of the other spins. 
Since $\langle 0_m|\calH_m|0_m \rangle$ = $E_{1}(0)$ 
does not involve ${\bf S}_{m}$ and ${\bf S}_{m+1}$, 
we have $\langle \Psi_{0}|H|\Psi_{0} \rangle = E_{1}(0)+E'$ 
with $E' = \langle \Psi' |H'|\Psi' \rangle$ 
being the ground-state energy of $H'$. 
Let $|\Psi \rangle$ be any state with $T_{m}=L \ge 1$. 
Then $\langle \Psi |H|\Psi \rangle = 
\langle \Psi |\calH_{m}|\Psi \rangle+\langle \Psi |H'|\Psi \rangle$. 
Clearly $\langle \Psi |\calH_{m}|\Psi \rangle \ge E_{1}(L)$ 
and $\langle \Psi|H'|\Psi \rangle\ge E'$;
also $E_{\rm 1}(L) > E_{\rm 1}(0)$ for 
$\lambda > \Lambda(L)$. 
Therefore 
$\langle \Psi |H|\Psi \rangle > \langle \Psi_{0}|H|\Psi_{0} \rangle$ 
and $|\Psi \rangle$ is not the ground state of $H$. 
\hfill $\square$ 

Since $\Lambda(L)$ decreases with $L$ 
from $\Lambda(1)=2S+1$ 
to $\Lambda(S)=2(2S+1)/(S+1)$, 
we obtain the following result: 

%=============================
{\bf Theorem 1.} 
If $\lambda > 2S+1$, then $T_{i}=0$ for all ${i}$ in the ground state. 
%=============================

\begin{figure}[t] %[!btp] [H]
\begin{center}\leavevmode
\includegraphics[width=1.00\linewidth]{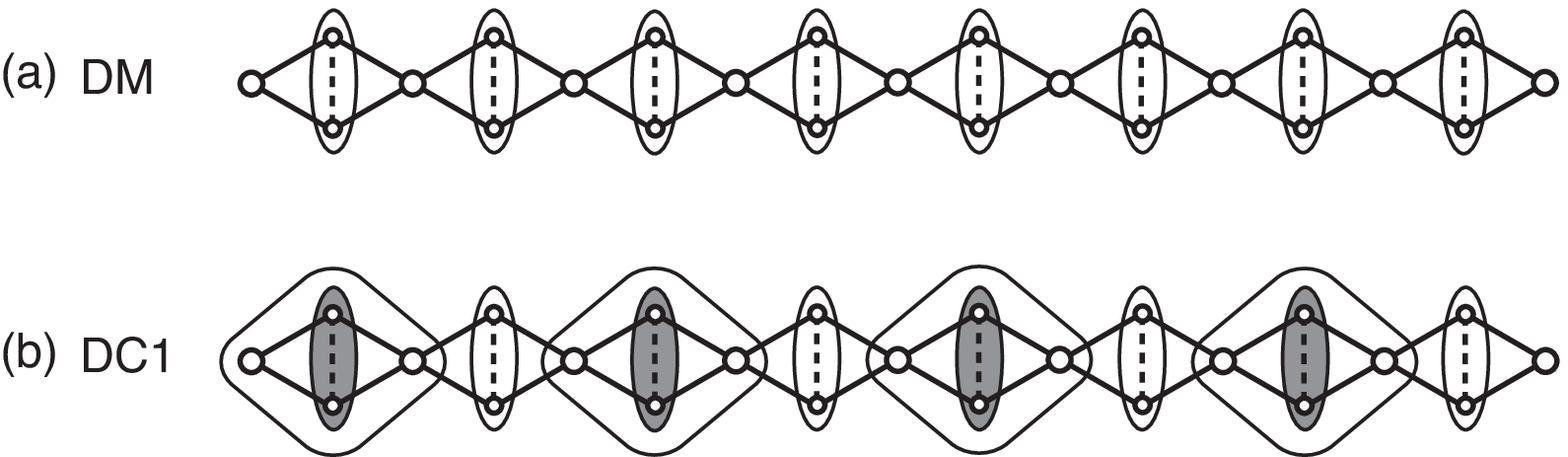}\\
\bigskip
\includegraphics[width=1.00\linewidth]{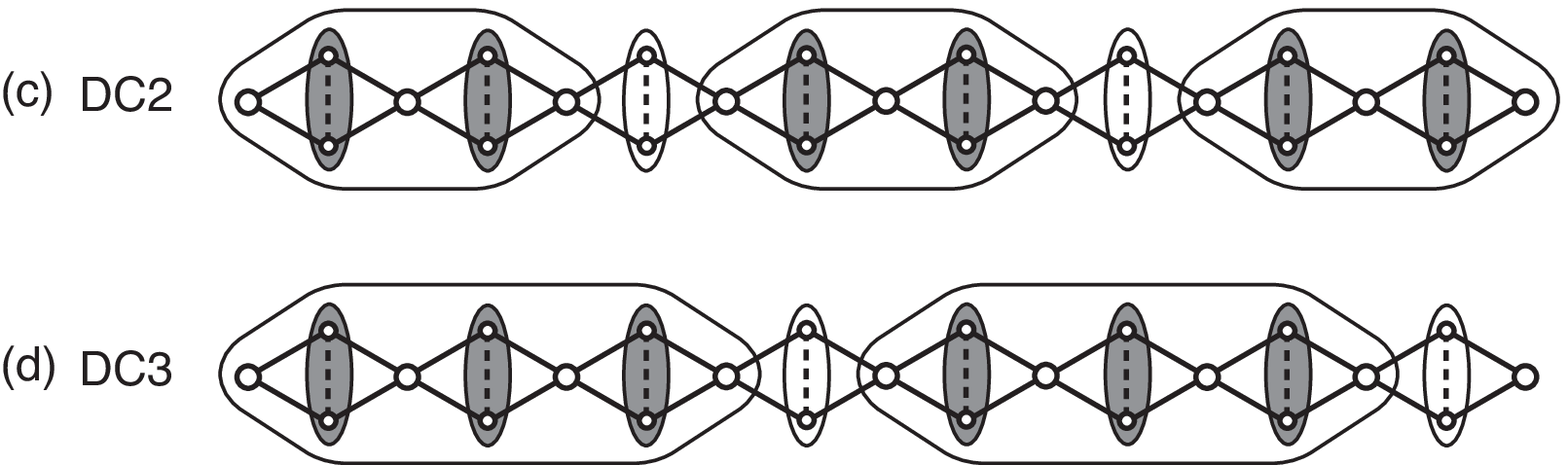}\\
\bigskip\smallskip
\includegraphics[width=1.00\linewidth]{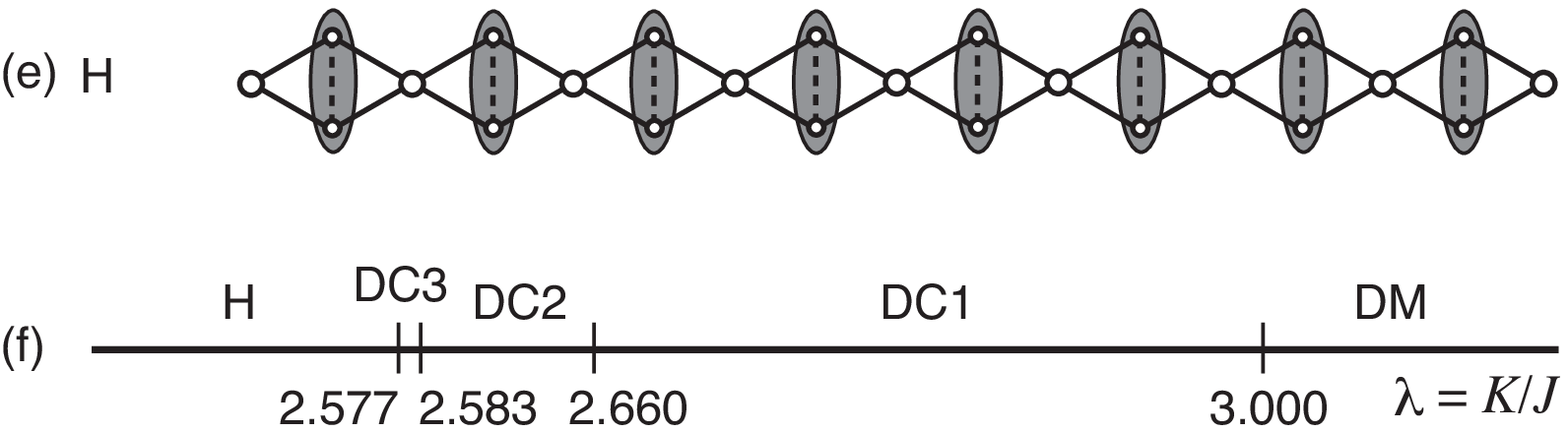}\\
\end{center}
\caption{The ground states for the MDC 
in the case of $S=1$ and $\tau=\frac{1}{2}$: 
(a) The dimer-monomer (DM) state. 
An unshaded oval represents a singlet dimer ($T_{i}$ = 0). 
There are free spins (monomers) on the sites not enclosed by ovals. 
(b) The dimer-cluster-1 (DC1) state. 
A shaded oval represents a triplet pair ($T_{i}$ = 1). 
A loop including a diamond unit is a cluster-1 
whose spin magnitude is unity. 
(c) The dimer-cluster-2 (DC2) state. 
A loop including two diamond unit is a cluster-2 
whose spin magnitude is unity. 
(d) The dimer-cluster-3 (DC3) state. 
A loop including a diamond unit is a cluster-3 
whose spin magnitude is unity. 
(e) The Haldane (H) state. 
(f) The ground-state phase diagram.} 
\label{state}
\end{figure}

In this ground state, all pairs of $\vv{\tau}_{i}^{(1)}$ and 
$\vv{\tau}_{i}^{(2)}$ form singlet dimers, so that 
all ${\bf S}_{i}$'s are decoupled from each other 
and behave as free spins.
Hence we call this state the {\it dimer-monomer} (DM) state. 
The DM state for $S=1$ is shown in Fig.~\ref{state}(a). 
The picture is also the same for the DM state of a system 
with $S \ge 2$. 
The ground state energy is $NE_1(0)$ for the lattice consisting of 
$N$ diamond units, where $2N$ spins with magnitude $\frac12 S$ 
and $N+1$ spins with magnitude $S$ are included. 
Due to the free spins, there is a $(2S+1)^{N}$-fold degeneracy 
in the DM state. 

%-->-->-->-->-->-->-->-->-->-->-->-->-->-->-->-->
%Fig. 2
%--<--<--<--<--<--<--<--<--<--<--<--<--<--<--<--<

For $\lambda < 2S+1$, we have the following lemma: 

%=============================
{\bf Lemma 2.} 
For $\lambda$ in $\frac{2}{3}(2S+1) < \lambda < 2S+1$, 
$T_{i}$ at any site is 0 or 1 in the ground state. 
Also at least one of $T_{i}$s is not 0. 
%=============================

{\bf Proof.} 
For $\tau = \frac{1}{2}$, the ground state is composed of 
$T_{i}$s with their magnitude 0 or 1. 
It is true even for $\tau \ge 1$ 
if $\Lambda(2) < \lambda < \Lambda(1)$ 
because of Lemma 1; 
the region is rewritten as $\frac{2}{3}(2S+1) < \lambda < 2S+1$. 
Further for $\lambda < 2S+1$, the DM state, where $T_{i} = 0$ 
for all $i$, is not a ground state, 
because we can lower the energy by introducing an isolated 
$T_{i}=1$ in the DM state as known from Eq.~(\ref{energy1}). 
\hfill $\square$ 

Hence $\lambda_{C1} \equiv \Lambda(1) = 2S+1$ 
is the phase boundary of the DM phase. 

For $\frac{2}{3}(2S+1) < \lambda < 2S+1$, the ground state 
consists of an appropriate array of $T_i = 0$ and 1. 
For further arguments, we define the {\it cluster}-$n$ as 
the local ground state of a successive $n$ diamond units 
in the case that $T_i$s are all unity in the $n$ diamond units 
and both $T_i$s just outsides of it are zero; 
the cluster-$n$ is a state of $n+1$ ${\bf S}_{i}$s, 
$n$ $\vv{\tau}_{i}^{(1)}$s and $n$ $\vv{\tau}_{i}^{(2)}$s. 
The magnetization of the cluster-$n$ is $(n+1)S-n$ 
due to the Lieb-Mattis theorem \cite{lieb-matt}. 
In particular, for $S=1$, the cluster-$n$ is the Haldane state 
of a finite Haldane chain with $2n+1$ sites, 
and is a triplet state. 
Denoting the energy of the cluster-$n$ by $E_{n}$, 
we define the energy per diamond unit 
for a cluster-$n$ measured from the DM state as 
%------------------------------------------------------------
\begin{align}
\label{e_n} %(7)
e_{n}(\lambda) = \frac{1}{n+1} [E_{n}-nE_{1}(0)]. 
\end{align}
%------------------------------------------------------------
We have divided it by $n+1$ and not by $n$, 
since a cluster-$n$ is accompanied with 
the neighboring $T_i = 0$s, and 
it is convenient to include the effect of one of them. 
If $e_{n}(\lambda)$ is the minimum at a single value of $n$, 
then the ground state in the thermodynamic limit 
($N\rightarrow \infty$) is realized by an alternating array of 
cluster-$n$s with $T_{i}=1$ and dimers with $T_{i}=0$. 
We call the state the {\it dimer-cluster}-$n$ (DC$n$) state. 
The DC$n$ states with $n$ = 1, 2, and 3 for $S=1$ are illustrated 
in Fig.~\ref{state}(b), (c), and (d), respectively. 
The DC$n$ state has spatial periodicity $n+1$, and 
the translational symmetry is spontaneously broken. 
The degeneracy is $(n+1)3^{N/(n+1)}$ for large $N$. 

To investigate the ground state of the MDC 
in the region of $T_i = 0$ or 1 for all $i$, we need 
the following lemma for small equivalent lattices: 

%=============================
{\bf Lemma 3.} 
$\tilE_{2} > 2\tilE_{1}$, 
if $T_i = 1$ for all $i$. 
%=============================

{\bf Proof.} 
We define three spin Hamiltonians 
$\tilH_{1L}=({\bf S}_1+{\bf S}_{2})\cdot{\bf T}_1$ 
and $\tilH_{1R}=({\bf S}_2+{\bf S}_{3})\cdot{\bf T}_2$. 
The five spin Hamiltonian for ${\bf S}_{1}$, ${\bf T}_{1}$, 
${\bf S}_{2}$, ${\bf T}_{2}$, and ${\bf S}_{3}$ is written as 
$\tilH_2$ = $\tilH_{1L}\otimes I_{T_2,S_3}+ I_{T_1,S_1}\otimes \tilH_{1R}$, 
where $I_{T,S}$ is an identity operator in the subspace of spins ${\bf T}$ and ${\bf S}$. 
The ground states of $\tilH_{1L}$, $\tilH_{1R}$, and $\tilH_2$ 
are denoted by $\ket{\psi_{3L}}$, $\ket{\psi_{3R}}$, 
and $\ket{\psi_{5}}$, respectively. 
From variational consideration, the following inequality holds: 
$\tilE_{2}=\bra{\psi_5}\tilH_2\ket{\psi_5} 
\geq \bra{\psi_{3L}}\tilH_{1L}\ket{\psi_{3L}}+\bra{\psi_{3R}}\tilH_{1R}\ket{\psi_{3R}}=2\tilE_{1}$. 

We now deduce a contradiction by assuming 
$\tilE_{2} = 2\tilE_{1}$.  
This is valid only if $\ket{\psi_{5}}$ is also the ground state of 
both $\tilH_{1L}\otimes I_{T_2,S_3}$ and 
$I_{T_1,S_1}\otimes \tilH_{1R}$. 
The Lieb-Mattis theorem \cite{lieb-matt} applied to $H_{1L}$ 
says that the magnitude of the total spin 
${\bf S}_1+{\bf T}_1+{\bf S}_2$ must be $2S-1$ in $\ket{\psi_{3L}}$. Therefore $\ket{\psi_{3L}}$ contains only the states with $S_1^z+T_1^z +S_2^z\leq 2S-1$. 
Applying the Lieb-Mattis theorem to $\tilH_2$, however, 
the ground state $\ket{\psi_5}$ has total spin $3S-2$ and 
contains all the $S^z$-diagonal states 
$\ket{S_1^z,T_1^z,S_2^z,T_2^z,S_3^z}$ 
with total spin $3S-2$. 
This implies that $\ket{\psi_5}$ contains the state with 
$S_1^z+T_1^z+S_2^z =2S$ such as $\ket{S, 0, S, -1, S-1}$ 
with finite amplitude. 
This is a contradiction and therefore we have 
$\tilE_{2} > 2\tilE_{1}$. 
\hfill $\square$ 

Based on Lemma 3, 
we have the following proposition on the ground state 
in a region just below $\lambda_{C1} = 2S+1$: 

%=============================
{\bf Theorem 2.} 
There exists a positive number $\delta$ such that 
the DC1 state is the ground state for 
$\lambda_{\rm C1} - \delta < \lambda < \lambda_{\rm C1}$ 
in the thermodynamic limit. 
%=============================

{\bf Proof.} 
Due to Lemma 2, we only have spin magnitudes 0 or 1 
for ${\bf T}_i$s for $\frac{2}{3}(2S+1) < \lambda < 2S+1$. 
We consider a segment of $n$ diamond units between two 0s. 
The equivalent lattice is described by 
Eq.~(\ref{equiv_Ham}) with $N=n$ and $T_i =1$ for all $i$. 
By using the ground state energy $\tilE_{n}$, 
we have $E_{n}-nE_{1}(0)=\tilE_{n}+n\lambda$. 
We adopt $\delta \equiv \tilE_{2} - 2\tilE_{1}$ $(>0)$ 
if it is smaller than $\frac{1}{3}(2S+1)$, and 
show that $\delta$ satisfies the condition of Theorem 2. 

If $n$ is even, the Hamiltonian for $\tilE_{n}$ is divided into 
$\frac{n}{2}$ sub-Hamiltonians, each equivalent to that for 
$\tilE_{2}$. 
A variational argument gives $\tilE_{n}$ $\ge$ 
$n\tilE_{2}/2$ = $n\tilE_{1}+n\delta /2$ 
for even $n$, and $\tilE_{n}$ $\ge$ 
$(n-1)\tilE_{2}/2 + \tilE_{1}$ = 
$n\tilE_{1}+(n-1)\delta /2$ for odd $n$. 
Since $e_{1}$ = $(\tilE_{1}+\lambda)/2$ 
= $(\lambda-\lambda_{\rm C1})/2$, 
a lower bound on $e_{n}-e_{1}$ is given as 
$\frac{n-1}{2(n+1)} (\lambda -\lambda_{\rm C1}+\delta)$, 
which is positive for $\lambda >\lambda_{\rm C1}-\delta$ 
and $n \ge 2$. 

If $\delta = \tilE_{2} - 2\tilE_{1} > \frac13 (2S+1)$, 
we use $\delta' \equiv \frac13 (2S+1)$ instead of $\delta$. 
Then $\lambda_{\rm C1}-\delta'$ is the lower bound 
of Lemma 2 where spin magnitudes of ${\bf T}_i$s are 0 or 1. 
The argument of the last paragraph stand still for $\delta'$ 
replacing $\delta$ since $0 < \delta' < \delta$. 
\hfill $\square$ 

For $\lambda \le 1$, we show that the ground state of the MDC 
is equivalent to that of 
the uniform linear spin chain with spin magnitude $S$. 

%=============================
{\bf Theorem 3.}  \ 
If $\lambda \le 1$, then $T_{i} = S$ for all $i$ in the ground state. 
%=============================

{\bf Proof.}
We divide the total lattice into two sublattices $A$ and $B$, 
where ${\bf S}_{i}$s are on the $A$ sublattice 
and $\vv{\tau}_{i}^{(\alpha)}$s on the $B$ sublattice. 
For $\lambda \le 1$, we have $K \le J$ 
for interaction $J$ $(>0)$ between spins on the different sublattices and 
interaction $K$ between spins on the same $B$ sublattice. 
This suits to the condition for the Lieb-Mattis 
theorem \cite{lieb-matt}. 
Hence the total spin of the ground state is given as 
$| (N+1)S - NS | = S$ for the lattice consisting of $N$ diamond units. 

We also apply the Lieb-Mattis theorem to 
the ground state of the equivalent 
lattice with $\{ T_{i}\}$. 
Then the total spin of the ground state is given as 
$| (N+1)S - \sum_{i=1}^{N} T_i |$. 
This must be the same value $S$ as the total spin of 
the ground state of the MDC. 
This is possible only if $T_{i} = S$ for all $i$. 
\hfill $\square$ 

Thus, at least for $\lambda \le 1$, 
the MDC is equivalent to a uniform linear chain 
with integer spin magnitude $S$. 
The ground state of an integer spin chain is 
the Haldane state \cite{haldane}. 
The picture of the Haldane state 
for the MDC with $S=1$ is shown in Fig.~\ref{state}(e). 

\begin{figure}[t] %[!btp] [H]
\begin{center}\leavevmode
\includegraphics[width=0.8\linewidth]{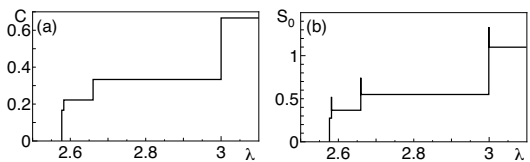}
\end{center}
\caption{$\lambda$-dependence of  the (a) Curie constant 
and (b) residual entropy for the MDC with $S=1$.}
\label{curie}
\end{figure}

For $S=1$, the DC$n$ ground state is resolved into 
the ground states of 
spin clusters equivalent to  finite-length spin-1 chains. 
Therefore we can precisely determine the phase boundaries 
by the exact numerical diagonalization for the finite chains. 
Equation (\ref{e_n}) becomes 
%------------------------------------------------------------
\begin{align} %(8)
e_n(\lambda) =\frac{\tilE_n-\lambda}{n+1}+ \lambda , 
\end{align}
%------------------------------------------------------------
where $\tilE_n$ is the ground state energy of the spin-1 chain 
with length $2n+1$.  
Typical values are $\tilE_0 = 0$, $\tilE_1 = -3$, 
$\tilE_2 = -5.8302125227708$, and 
$\tilE_3 = -8.6345319827062$. 
If  $T_i=1$ for all $i$, the ground state energy per unit cell 
is given by 
$e_\infty(\lambda)$ $=2\tilde{\epsilon}_{\infty} +\lambda$, 
where $\tilde{\epsilon}_{\infty} \simeq -1.401484038971$ \cite{white}
 is the ground state energy of an infinite spin-1 chain per unit cell. 
The phase transition between the DC$(n-1)$ and DC$n$ phases takes place at $\lambda=\lambda_{{\rm C}n}$ 
$\equiv (n+1)\tilE_{n-1} - n\tilE_n$, which is the solution of 
$e_{n-1}(\lambda_{{\rm C}n})=$ $e_n(\lambda_{{\rm C}n})$.  
However the DC$n$ phases with $n \ge 4$ 
do not appear \cite{ncheck} and 
a phase transition takes place directly from the DC3 phase 
to the Haldane (DC$\infty$) phase at 
$\lambda_{\rm C\infty}=\tilE_3 - 8\tilde{\epsilon}_{\infty}$, 
which is the solution of 
$e_3(\lambda_{\rm C\infty})=$ 
$e_{\infty}(\lambda_{\rm C\infty})$. 
The critical values of $\lambda$ are estimated as 
$\lambda_{\rm C1} = 2S+1 = 3$, 
$\lambda_{\rm C2} = 2.660425045542$, 
$\lambda_{\rm C3} = 2.58274585704$, and 
$\lambda_{\rm C\infty} = 2.5773403291$. 
Thus we have the ground-state phase diagram for $S=1$ 
in Fig.~\ref{state}(f). 
Magnetically, each phase is characterized by 
the number of alive spins; it equals the number $N/(n+1)$ of 
triplet clusters in the DC$n$ ground state. 
Consequently, as an experimentally measurable quantity, 
the Curie constant  given by $C =\frac{2}{3}(n+1)^{-1}$ 
shows a stepwise $\lambda$-dependence 
as shown in Fig.~\ref{curie}(a). 
Also, the residual entropy $S_0$ per unit length 
is $\ln 3/(n+1)$ in the DC$n$ phase. 
At the phase boundary between the DC$n$ and DC$(n+1)$ phases, 
the mixing of cluster-$n$ and cluster-$(n+1)$ 
remarkably enhances the entropy, 
which is estimated by combinatory argument as 1.333, 0.744, and 0.522 for $n =$ 0, 1, and 2, respectively. 
Thus $S_0$ has spike-like structures 
as shown in Fig.~\ref{curie}(b). 
Similar structures are also reported in the diamond hierarchical 
Ising model~\cite{fuku}. 

%-->-->-->-->-->-->-->-->-->-->-->-->-->-->-->-->
%Fig. 3
%-->-->-->-->-->-->-->-->-->-->-->-->-->-->-->-->

Summarizing, we investigated the MDC and 
proved that the DM, DC1, and Haldane phases exist 
for an arbitrary integer spin magnitude $S$. 
For $S=1$, we numerically determined 
all the phase boundaries of the full phase diagram, 
which consists of the DM, DC1, DC2, DC3, and Haldane phases.  
Since each spin cluster for magnitude $S$ has 
$2S+1$ degrees of freedom, 
the total ground state is massively degenerate. 
If this degeneracy is lifted by an appropriate perturbation, 
a variety of exotic phases are expected. 

KT and HS are supported by the
Fund for Project Research of Toyota Technological Institute. 
KH is  supported by a Grant-in-Aid for Scientific Research  on Priority Areas, "Novel States of Matter Induced by Frustration" (20046003), from the Ministry of Education, Science, Sports and Culture of Japan.
He also thanks 
the Supercomputer Center, Institute for Solid State Physics, 
University of Tokyo and Supercomputing Division, Information Technology Center, University of Tokyo
for the use of  the facilities. 
Part of the numerical calculation is carried out by 
using the program based on the package TITPACK ver.2 
coded by H. Nishimori.


\begin{references}
\bibitem{ncheck} %[13]
This has been checked numerically up to $n=9$.
\end{references}
\end{document}